\documentclass[prl,twocolumn,superscriptaddress]{revtex4}
\usepackage{graphicx}
\usepackage{subfigure}
\usepackage{amsmath}

\usepackage{color} 

\newcommand{\ket}[1]{| #1 \rangle}
\newcommand{\bra}[1]{\langle #1 |}
\newcommand{\ketbra}[2]{\ket{#1}\!\bra{#2}}

\begin{document}

\title{Measurement of Autler-Townes and Mollow transitions\\
in a strongly driven superconducting qubit
}

\author{M.~Baur}
\author{S.~Filipp}
\author{R.~Bianchetti}
\author{J.M.~Fink}
\author{M.~G\"oppl}
\author{L.~Steffen}
\author{P.J.~Leek}
\affiliation{Department of Physics, ETH Zurich, CH-8093 Zurich,
Switzerland}

\author{A.~Blais}
\affiliation{D\'epartement de Physique, Universit\'e de Sherbrooke,
Sherbrooke, Qu\'ebec J1K2R1, Canada}

\author{A.~Wallraff}
\affiliation{Department of Physics, ETH Zurich, CH-8093 Zurich,
Switzerland}

\date{\today}
\begin{abstract}
We present spectroscopic measurements of the Autler-Townes doublet
and the sidebands of the Mollow triplet in a driven superconducting
qubit. The ground to first excited state transition of the qubit is
strongly pumped while the resulting dressed qubit spectrum is probed
with a weak tone. The corresponding transitions are detected using
dispersive read-out of the qubit coupled off-resonantly to a
microwave transmission line resonator. The observed frequencies of
the Autler-Townes and Mollow spectral lines are in good agreement
with a dispersive Jaynes-Cummings model taking into account higher
excited qubit states and dispersive level shifts due to off-resonant
drives.
\end{abstract}

\maketitle

When a two-level system is driven on resonance with a strong
monochromatic field, the excited state population undergoes coherent
Rabi oscillations. This coherent process is reflected in the
appearance of two sidebands offset by the Rabi frequency from the
main qubit transition in the spectrum. This leads to a three peaked
fluorescence spectrum referred to as the Mollow triplet
\cite{Mollow1969}. When probing transitions into a third atomic
level, two characteristic spectroscopic lines separated by the Rabi
frequency appear, a feature which is called the Autler-Townes
doublet \cite{Autler1955}. The Mollow triplet and the Autler-Townes
doublet were observed for the first time in an atomic beam of sodium
\cite{Wu1975} and in a He-Ne discharge laser \cite{Schabert1975},
respectively. Later they have been measured in single molecules
\cite{Wrigge2008,Tamarat1995}, single atoms \cite{Walker1995} and
more recently also in quantum dots
\cite{Xu2007,Vamivakas2008,Muller2007}.

Here we present experiments in which we spectroscopically probe Mollow
sideband and Autler-Townes transitions in a strongly driven
superconducting quantum electronic circuit with discrete energy
levels \cite{Clarke2008}. The properties of superconducting qubits
dressed by strong drive fields have also been studied
experimentally in Refs.~\onlinecite{Wilson2007,Oliver2005}. Other examples of
spectroscopic techniques used in the context of superconducting
qubits include multi-photon spectroscopy with photons of the same
\cite{Wallraff2003,Bishop2008} and of different frequencies \cite{Saito2006}, amplitude
spectroscopy \cite{Berns2008}, side-band spectroscopy of coupled
systems \cite{Wallraff2007} and pump/probe spectroscopy \cite{Fink2008}. In several experiments it has also been shown that
artifical atoms based on superconducting circuits show quantum
optical effects as real atoms do \cite{Schoelkopf2008}. Single
photons \cite{Houck2007}, Fock states generation \cite{Hofheinz2008}
and lasing effects in a Cooper pair box \cite{Astafiev2007} have
been demonstrated.

\begin{figure}
  \centering
  \includegraphics[width=\columnwidth]{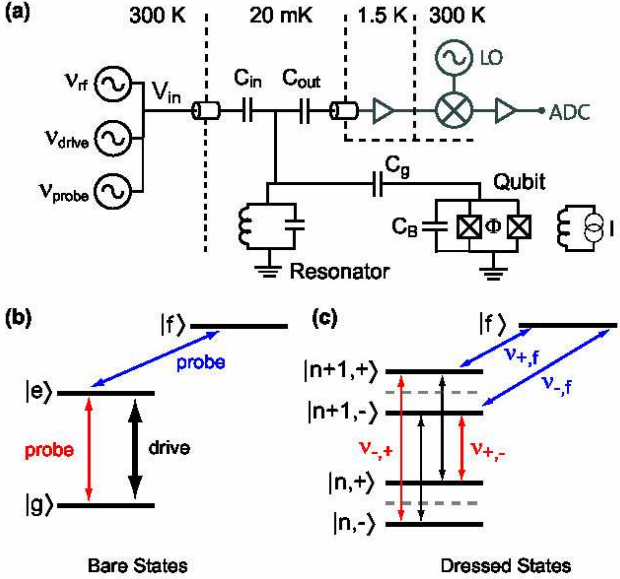}
\caption{ (a) Simplified circuit diagram of the measurement setup
analogous to the one used in Ref.~\cite{Fink2008}. In the center at
the 20 mK stage, the qubit is coupled capacitively through $C_g$ to
the resonator, represented by a parallel LC oscillator, and the
resonator is coupled to the input and output transmission lines over
capacitances $C_\mathrm{in}$ and $C_\mathrm{out}$. Three microwave
signal generators are used to apply the measurement
$\nu_{\mathrm{rf}}$ and drive and probe tones
$\nu_{\mathrm{drive/probe}}$ to the input port of the resonator. The
transmitted measurement signal is then amplified by an ultra-low
noise amplifier at 1.5 K, down-converted with an IQ-mixer and a
local oscillator (LO) to an intermediate frequency at 300K and
digitized with an analog-to-digital converter (ADC). (b)
Energy-level diagram of a bare three-level system with states
$\ket{g},\ket{e},\ket{f}$ ordered with increasing energy. Drive and probe
transitions are indicated by black and red/blue arrows, respectively. (c)
Energy-level diagram of the dipole coupled dressed states with the
coherent drive tone. Possible transitions induced by the probe tone
between the dressed states and the third qubit level
($\nu_{-,f}$,$\nu_{+,f}$) and between the dressed states
($\nu_{-,+},\nu_{+,-}$) are indicated with blue and red arrows. }
  \label{fig:fig1}
\end{figure}

In the experiments presented here, we use a version of the Cooper
pair box \cite{Bouchiat1998}, called transmon qubit \cite{Koch2007},
as our multilevel quantum system. States of increasing energies are
labelled $\ket{l}$ with $l=g,e,f,h,i,\ldots$ The transition
frequency $\omega_{ge}$ between the ground $\ket{g}$ and first excited state
$\ket{e}$ is approximated by $\hbar\omega_{ge}\approx\sqrt{8
E_CE^{\rm{max}}_J|\cos{2\pi\Phi/\Phi_0}|}-E_C$ \cite{Koch2007},
where $E_C/h = 233\,\rm{MHz}$ is the charging energy and
$E_J^{\mathrm{max}}/h =32.8\,\rm{GHz}$ is the maximum Josephson
energy. The transition frequency $\omega_{ge}$ can be controlled by
an external magnetic flux $\Phi$ applied to the SQUID loop formed by
the two Josephson junctions of the qubit. The transition frequency
from the first $\ket{e}$ to the second excited state $\ket{f}$ is
given by $\omega_{ef}=\omega_{ge}-\alpha$, where $\alpha\approx 2\pi E_C/h$
is the qubit anharmonicity \cite{Koch2007}. The qubit is strongly
coupled to a coplanar waveguide resonator with resonance frequency
$\omega_r/2\pi = 6.439\,\rm{GHz}$ and photon decay rate
$\kappa/2\pi\approx 1.6$ MHz. A schematic circuit diagram of the
setup is shown in Fig~\ref{fig:fig1}(a).

When the ground to first excited state transition of the qubit is in
resonance with the resonator
($\Delta_{ge}=\omega_{ge}-\omega_{r}=0$), the strong coupling gives
rise to the vacuum Rabi mode splitting \cite{Wallraff2004,Fink2008,Bishop2008}
from which we have determined a dipole coupling strength $g_{ge}/2\pi=133\,\rm{MHz}$ between the
first two energy levels. In the
non-resonant regime, where the qubit is far detuned from the
resonator ($|\Delta_{ge}|\gg g_{ge}$), the system is described by
the generalized Jaynes-Cummings Hamiltonian in the dispersive limit
\cite{Koch2007}
\begin{align}
  \label{eq:4}
  H_{JC}\approx & \hbar\Bigl [\omega_r -\chi_{ge}\ketbra{g}{g}
  +\sum_{l=e,f,\ldots}(\chi_{l-1,l}-\chi_{l,l+1})\ketbra{l}{l}\Bigr ]a^\dagger a \notag \\
  &+ \hbar\omega_{g}\ketbra{g}{g}+\hbar \sum_{l=e,f,\ldots} (\omega_{l}+\chi_{l-1,l})\ketbra{l}{l}.
\end{align}
In the first term, $\chi_{ge}$ and $(\chi_{l-1,l}-\chi_{l,l+1})$
describe both the qubit-state dependent resonator frequency shift
and the ac-Stark shift of the qubit energy levels
\cite{Schuster2005,Schuster2007,Koch2007}. The dispersive frequency
shift $\chi_{l,l+1}=g^2_{l,l+1}/\Delta_{l,l+1}$ is determined by the
coupling strength $g_{l,l+1}$ between the levels $\ket{l}$ and
$\ket{l+1}$ mediated by the resonator field and the detuning
frequency $\Delta_{l,l+1}=\omega_{l,l+1}-\omega_r$. $a$ ($a^\dagger$)
are the annihilation (creation) operators of the single mode field.
In the last term, $\chi_{l-1,l}$ describes the Lamb shift of the
transmon levels due to the dispersive coupling of the qubit to
vacuum fluctuations in the resonator \cite{Fragner2008}.

\begin{figure}
  \includegraphics[width=\columnwidth]{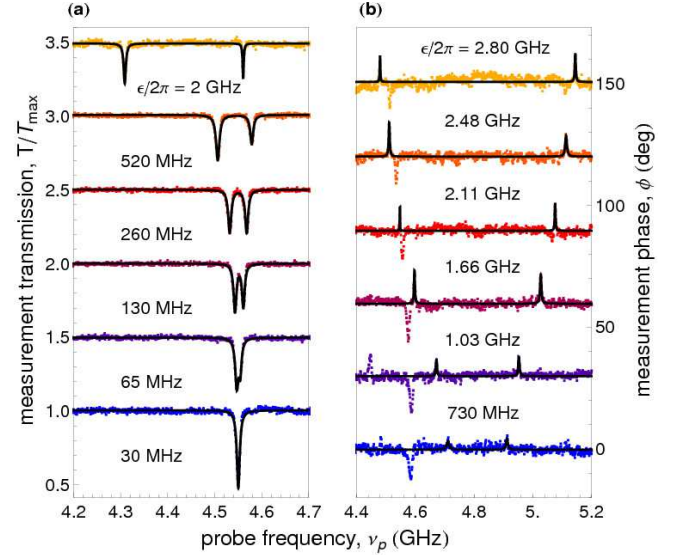}
  \caption{\label{fig:fig2}
    (a) Autler-Townes spectrum as a function of
    drive amplitude $\varepsilon$. Traces are normalized to the maximum
    transmission through the resonator, and separated from each other
    with a vertical offset of 0.5.
    (b) Mollow spectrum in phase.
    Traces are offset by 30 degrees. Black solid lines are fits to
    Lorentzians. Peaks not fitted with Lorentzians correspond to the phase
    response of the Autler-Townes doublet.
  }
\end{figure}

We measure the Autler-Townes and the Mollow spectral lines according
to the scheme shown in Fig.~\ref{fig:fig1}(b). First, we tune the
qubit to the frequency $\omega_{ge}/2\pi \approx 4.811\,\rm{GHz}$, where it
is strongly detuned from the resonator by $\Delta /2\pi =
1.63\,\rm{GHz}$. We then strongly drive the transition
$\ket{g}\rightarrow\ket{e}$ with a first microwave tone of amplitude
$\varepsilon$ applied to the qubit at the fixed frequency
$\omega_d=4.812\,\rm{GHz}$. The drive field is described by the
Hamiltonian $H_{d}=\hbar\varepsilon (a^\dagger e^{-i\omega_dt}+a
e^{i\omega_d t})$ where the drive amplitude $\varepsilon$ is given
in units of a frequency. The qubit spectrum is then probed by
sweeping a weak second microwave signal over a wide range of
frequencies $ \omega_p$ including $\omega_{ge}$ and $\omega_{ef}$.
Simultaneously, amplitude $T$ and phase $\phi$ of a
microwave signal applied to the resonator are measured
\cite{Wallraff2004}. We have adjusted the measurement frequency to
the qubit state-dependent resonance of the resonator under qubit
driving for every value of $\varepsilon$.
Figures~\ref{fig:fig2}(a) and (b) show the measurement response $T$
and $\phi$ for selected values of $\varepsilon$. For drive
amplitudes $\varepsilon/2\pi > 65\,\rm{MHz}$, two peaks emerge in
amplitude from the single Lorentzian line at frequency $\omega_{ef}$
corresponding to the Autler-Townes doublet, see Fig.~\ref{fig:fig2}(a).
The signal corresponding to the sidebands of the Mollow triplet is visible at high
drive amplitudes $\varepsilon/2\pi > 730\,\rm{MHz}$ in phase, see
Fig.~\ref{fig:fig2}(b). Black lines in Fig.~\ref{fig:fig2} are fits
of the data to Lorentzians from which the dressed qubit resonance
frequencies are extracted.

An intuitive model explaining those two effects can be given in the
dressed state picture \cite{Cohen-Tannoudji1998}. In the situation
where the drive is exactly on resonance with the qubit, the bare
states $\ket{n,g}$ and $\ket{n-1,e}$ of the uncoupled atom-field
system are degenerate, where $n$ is the average number of photons in
the coherent drive. The dipole coupling splits the energy levels by
the Rabi frequency $\hbar\Omega_R$ and forms an energy ladder of
doublets separated by the energy of the drive photons $\hbar
\omega_d$. The new dressed eigenstates dipole coupled to the field are
symmetric and antisymmetric superpositions of the bare states
$\ket{n,\pm}=\ket{n,g}\pm\ket{n-1,e}$, see Fig.~\ref{fig:fig1}(c).
In the limit $n \gg \sqrt{n}$, the allowed transitions between dressed state doublets
appear at frequencies $\omega_{0}=\omega_{ge}$,
$\omega_{+,-}=\omega_{ge}-\Omega_R$ and
$\omega_{-,+}=\omega_{ge}+\Omega_R$ which are the central line and
the two sidebands of the Mollow triplet, respectively, indicated by
black and red arrows in Fig.~\ref{fig:fig1}c. The central line is not observed in our measurements as the corresponding transition is completely saturated by the strong drive tone. Similarly, transitions
from one pair of dressed levels $\ket{n+1,\pm}$ to the third level
$\ket{f}$ at frequencies $\omega_{\pm,f}=\omega_{ef}\mp\Omega_R/2$
correspond to the Autler-Townes doublet. The splitting of the
dressed states is only well resolved, when $\Omega_{R}$ is
considerably larger than the qubit linewidth.

\begin{figure}
\includegraphics[width=\columnwidth]{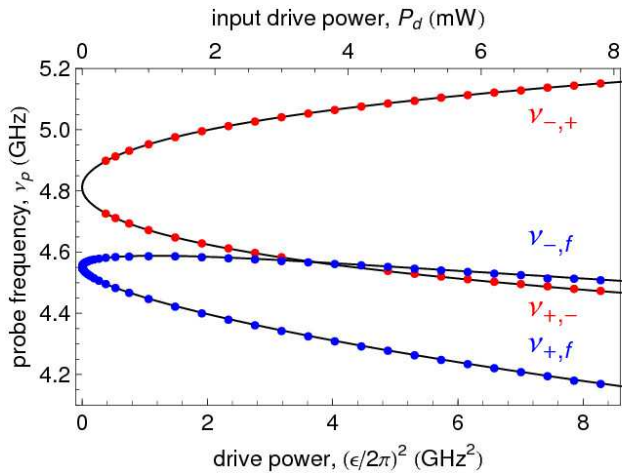}
\caption{\label{fig:fig3} Measured Autler-Townes doublet (blue dots)
and Mollow triplet sideband frequencies (red dots) \textsl{vs.} drive power
$P_d$ at a fixed drive frequency $\omega_d/2\pi=4.812\,\rm{GHz}$. Black
solid lines are transition frequencies calculated by numerically
diagonalizing the Hamiltonian (\ref{eq:2}) taking into account the lowest 5
transmon levels.}
\end{figure}

The frequencies of the Autler-Townes doublet (blue data points) and
of the Mollow triplet sidebands (red data points) extracted from the
Lorentzian fits in Fig.~\ref{fig:fig2}(a) and (b) are plotted in
Fig.~\ref{fig:fig3}. The splitting of the spectral lines in pairs
separated by $\Omega_R$ and $2 \Omega_R$, respectively, is observed
for Rabi frequencies up to $\Omega_R/2\pi \approx 300 \, \rm{MHz}$
corresponding to about 6\% of the qubit transition frequency
$\omega_{ge}$.

In the simplest model, the continuous classical drive at frequency
$\omega_d$ is expected to induce Rabi oscillations between the qubit
levels $\ket{l}$ and $\ket{l+1}$ at the frequency \cite{Blais2007}
\begin{equation}
  \label{eq:1}
  \Omega_{l,l+1}\approx\dfrac{2\varepsilon g_{l,l+1}}{\omega_r -\omega_d},
\end{equation}
depending linearly on the drive amplitude $\varepsilon$. Therefore,
one would expect that the strong drive at the qubit transition
frequency $\omega_d\approx\omega_{ge}$ should lead to a square-root
dependence of the Autler-Townes and Mollow spectral lines on the
drive power $P_d\propto \varepsilon ^2$. However, the Autler-Townes
spectral lines show a clear power dependent shift, see
Fig.~\ref{fig:fig3}, and the splitting of both pairs of lines scales
weaker than linearly with $\varepsilon$.

These effects can be fully understood calculating the
different transition frequencies by numerically diagonalizing the
Jaynes-Cummings Hamiltonian in the dispersive limit including
the coherent drive on the qubit
\begin{equation}
  \label{eq:2}
  H\approx H_{JC}+\sum_l \dfrac{\Omega_{l,l+1}}{2}
  (\ketbra{l}{l+1}e^{i\omega_d t}+h.c.).
\end{equation}
Here we take into account only the drive terms between nearest
neighbor energy levels since other transitions are strongly
suppressed due to the near harmonicity of the transmon
\cite{Koch2007}. This model is in good agreement with our data when
considering the lowest 5 qubit levels, see solid black lines in
Fig.~\ref{fig:fig3}. Because of the low
anharmonicity~\cite{Koch2007} and large drive amplitude, many qubit
levels must be included in the description. The calibration factor
between the externally applied drive amplitude and $\varepsilon$ is
the only free parameter in the fit.

Numerical diagonalization of Eq.~(\ref{eq:2}) also leads to a
qualitative understanding of the amplitude and phase
information contained in the measurement signal.
This is done by first calculating the pulled cavity frequencies using the pre-factor
of $a^\dag a$ in $H_{JC}$. Since the measurement rate is
small~\cite{Gambetta2008}, the measured signal is given by the
averaged response of all the dressed-state pulled frequencies
contained in the steady-state reached by the qubit under the strong drive
tone. In the Autler-Townes configuration, the weak probe tone
transfers a small population from the dressed ground and excited
states to the dressed $f$ state, resulting in a change in the cavity frequency and a 
drop of transmitted
signal. On the other hand, in the Mollow configuration, population
is exchanged by the probe tone from the $g$ to the $e$
dressed-states.  At low drive power, the dressed $g$ and $e$ states
are equal superpositions of the bare $g$ and $e$ states such that no
signal is measured. As the power is increased, these states get
dressed in different proportion with $f$ and a signal is measured.

\begin{figure}
  \centering
  \includegraphics[width=\columnwidth]{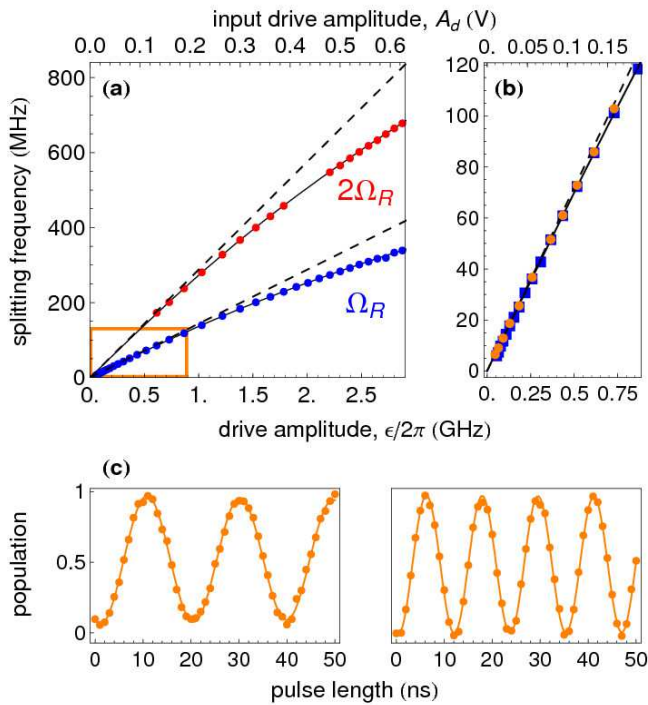}
\caption{(a) Extracted splitting frequencies of the Mollow triplet
sidebands (red dots) and the Autler-Townes doublet (blue dots) as a
function of the drive field amplitude. Dashed lines: Rabi
frequencies obtained with Eq.~(\ref{eq:1}). Black solid lines:
Rabi frequencies calculated by numerically diagonalizing the
Hamiltonian Eq.~(\ref{eq:2}) taking into account 5 transmon
levels. (b) Zoom in of the region in the orange rectangle in (a).
Orange dots: Rabi frequency $\Omega_{ge}$ \textsl{vs.} drive
amplitude $\varepsilon$ extracted from time resolved Rabi
oscillation experiments, lines as in (a). (c) Rabi oscillation
measurements between states $\ket{g}$ and $\ket{e}$ with $\Omega_R/2\pi = 50\,\rm{MHz}$ and $85\,\rm{MHz}$. }
  \label{fig:fig4}
\end{figure}

Finally, plotting the difference between the two Autler-Townes
spectral lines (blue data points) and the sidebands of the Mollow
spectrum (red data points) versus drive amplitude $\varepsilon$, the
nonlinearity of the dressed state splitting becomes more apparent,
see Fig.~\ref{fig:fig4}(a). The dashed line shows the linear
dependence of the Rabi frequency Eq.~(\ref{eq:1}) on the drive
amplitude $\varepsilon$, which only fits to the data at low
$\varepsilon$. The non-linear dependence at high $\varepsilon$,
instead, agrees very well only with our full model, black solid
line.

To confirm the direct relationship between the measured dressed
state splitting frequency and the Rabi oscillation frequency of the
excited state population we have also performed time resolved
measurements of the Rabi frequency up to $100\,\rm{MHz}$, see
Fig.~\ref{fig:fig4}(c). The extracted Rabi frequencies (orange data
points) are in good agreement with the spectroscopically measured
Rabi frequencies (blue squares) over the range of accessible
$\varepsilon$, as shown in Fig.~\ref{fig:fig4}(b)

In conclusion, we have observed the dressed state splitting of the
strongly driven energy levels of a superconducting qubit. The
frequencies of the Autler-Townes doublet and sidebands of the Mollow
triplet determined using a dispersive measurement technique are in
excellent agreement with theory. Splittings corresponding to Rabi
frequencies of up to $300\,\rm{MHz}$ have been observed
spectroscopically and are consistent with time resolved
measurements. Dressed state splittings have also been suggested to
realize tunable coupling between two qubits biased at their optimal
points \cite{Rigetti2005}. Our measurements are the first step
towards the realization of this protocol.

\begin{acknowledgments}
We thank Maxime Boissonneault and Jay Gambetta for useful
discussions. This work was supported by the Swiss National Science
Foundation and by ETH Z\"urich. P.J.L. acknowledges support from the
EC via an Intra-European Marie-Curie Fellowship. A.B. was supported
by NSERC, CIFAR, FQRNT and Alfred P. Sloan Foundation.
\end{acknowledgments}

\bibliographystyle{apsrev}

\end{document}